\documentclass[twocolumn,aps,pra,floatfix]{revtex4-1}
\usepackage{graphicx}
\usepackage{times}
\usepackage{nicefrac}
\usepackage{amsmath}
\usepackage{amsfonts}
\usepackage{amssymb}
\usepackage{amsthm}
\usepackage{epsf}
\usepackage{bm}
\usepackage{times}
\allowdisplaybreaks

\usepackage{dcolumn}
\usepackage{siunitx}

\newcommand{\veps}{\varepsilon}
\newcommand{\balpha}{\bm{\alpha}}

\newcommand{\bfp}{{\bf p}}
\newcommand{\bfA}{{\bf A}}
\newcommand{\bfD}{{\bf D}}
\newcommand{\bfr}{{\bf r}}

\newcommand{\bfnr}{{\bf{\hat r}}}

\newcommand{\be}{\begin{eqnarray}}
\newcommand{\ee}{\end{eqnarray}}

\newcommand{\dvec}[2]{{\left( \begin{array}{c} #1 \\ #2 \\ \end{array} \right)}}

\newcommand{\la}{\langle}
\newcommand{\ra}{\rangle}

\usepackage{color}
\definecolor{BLUE}{rgb}{0.0,0.0,1.0}

\usepackage{soul}
\soulregister\cite7
\soulregister\ref7

\begin{document}

\title{QED calculations of the nuclear recoil effect on the bound-electron $g$ factor}

\author{A. V. Malyshev}
\affiliation {Department of Physics, St.~Petersburg State University, 
Universitetskaya 7/9, 199034 St.~Petersburg, Russia}

\author{D. A. Glazov}
\affiliation {Department of Physics, St.~Petersburg State University, 
Universitetskaya 7/9, 199034 St.~Petersburg, Russia}

\author{V. M. Shabaev}
\affiliation {Department of Physics, St.~Petersburg State University, 
Universitetskaya 7/9, 199034 St.~Petersburg, Russia}

\begin{abstract}

Fully relativistic approach is applied to evaluation of the nuclear recoil effect on the bound-electron $g$ factor in hydrogenlike ions to first order in the electron-to-nucleus mass ratio~$m/M$ and to all orders in $\alpha Z$. The calculations are performed in the range $1 \leqslant Z \leqslant 20$ for the $g$ factors of $1s$, $2s$, $2p_{1/2}$, and $2p_{3/2}$ states. The $\alpha Z$-dependence of the nontrivial QED recoil contribution as a function of $Z$ is studied. 

\end{abstract}

\maketitle

\section{Introduction}

In recent decades, considerable progress in theoretical and experimental investigations of the bound-electron $g$ factor in few-electron ions has been achieved (for review, see, e.g., Refs.~\cite{Shabaev:2015:031205, Sturm:2017:4} and references therein). For instance, high-precision measurements of the $g$ factor in hydrogenlike ions accompanied with the elaborate quantum electrodynamics (QED) calculations lead to the most accurate determination of the electron mass~\cite{Haffner:2000:5308, Verdu:2004:093002, Sturm:2011:023002, Sturm:2013:030501_R, Sturm:2014:467, Zatorski:2017:012502}. On the other hand, comparing experimental data and theoretical predictions provides the most stringent test of the magnetic sector of bound-state QED to date. The $g$-factor investigations in lithiumlike \cite{Wagner:2013:033003, Volotka:2014:253004, Kohler:2016:10246, Yerokhin:2017:062511, Glazov:2019:173001} and boronlike \cite{Arapoglou:2019:253001} ions open possibilities to study the many-electron QED effects on the Zeeman splitting. There are also proposals how to employ these studies for an independent determination of the fine-structure constant~\cite{Shabaev:2006:253002, Volotka:2014:023002, Yerokhin:2016:100801}.

The measurement of the isotope shift of the ground-state $g$ factor in Li-like calcium~\cite{Kohler:2016:10246} has triggered a special interest to the relativistic calculations of the $g$-factor contribution due to the nuclear recoil effect. The fully relativistic description of this effect on the atomic $g$ factor requires the development of QED approaches which are beyond the usual Furry picture formalism~\cite{Furry:1951:115}, i.e., beyond the external-field approximation which treats the nucleus merely as a source of the classical electromagnetic field. The first fully relativistic evaluation of the recoil contribution to the $1s$ $g$ factor was performed in Ref.~\cite{Shabaev:2002:091801} using the QED formalism developed in Ref.~\cite{Shabaev:2001:052104}. In Ref.~\cite{Shabaev:2017:263001}, the effective four-component operators to treat the nuclear recoil effect on the atomic $g$ factor within the lowest-order relativistic (Breit) approximation were derived. With the help of these operators, the most precise theoretical predictions for the nuclear recoil contribution to the bound-electron $g$ factor in lithiumlike ions were obtained~\cite{Shabaev:2017:263001, Shabaev:2018:032512}. A possibility to probe the fully relativistic QED recoil contribution on a few-percent level in a specific difference of the $g$ factors of heavy H- and Li-like ions was discussed in Ref.~\cite{Malyshev:2017:765}. Finally, the nuclear recoil contribution to the bound-electron $g$ factor in B-like ions was considered in Refs.~\cite{Shchepetnov:2015:012001, Glazov:2018:457, Aleksandrov:2018:062521}.

The present study is devoted to the high-precision QED evaluation of the nuclear recoil effect on the bound-electron $g$ factor of the $1s$, $2s$, $2p_{1/2}$, and $2p_{3/2}$ states in H-like ions in the range $Z=1-20$. For the $s$ states, the  previous calculations of the QED recoil contribution to the $g$ factor are extended in order to cover all the ions within the range specified. For particular ions which were considered previously~\cite{Shabaev:2002:091801, Shabaev:2017:263001}, the accuracy of the theoretical predictions is improved. For the $2p_{1/2}$ state, to date this term was evaluated for $Z \geqslant 20$ only~\cite{Aleksandrov:2018:062521}. The QED recoil contribution to the $g$ factor of the $2p_{3/2}$ state has not been yet considered. The $\alpha Z$-dependence of all the obtained values is studied and the leading orders in $\alpha Z$ are extracted. The nuclear recoil effect on the $g$ factor of few-electron ions comprises the one-electron contribution evaluated in the present work and the many-electron contributions which can be calculated within the Breit approximation employing the corresponding effective operators \cite{Shabaev:2017:263001}. These calculations are in demand in view of the presently implemented ARTEMIS experiment~\cite{Lindenfels:2013:023412, Vogel:2019:1800211} at GSI in Darmstadt and ALPHATRAP experiment at the Max-Planck-Institut f\"ur Kernphysik (MPIK) in Heidelberg~\cite{Arapoglou:2019:253001, Sturm:2019:1425}. These experiments are expected to attain the accuracy of $10^{-9}-10^{-10}$ and better for the $g$ factors of low- and high-$Z$ few-electron ions~\cite{Sturm:2017:4}. Therefore, the proper treatment of the nuclear recoil effect on the bound-electron $g$ factor is an urgent task.

Relativistic units ($\hbar=1, c=1$) and Heaviside charge unit ($e^2=4\pi\alpha$, $e<0$) are employed throughout the paper.

\section{Theoretical methods}

The fully relativistic theory of the nuclear recoil effect on the bound-electron $g$ factor to first order in the electron-to-nucleus mass ratio $m/M$ and to all orders in $\alpha Z$ ($\alpha$ is the fine-structure constant and $Z$ is the nuclear charge number) is formulated in Ref.~\cite{Shabaev:2001:052104}. Let us briefly review the basic results obtained therein for a hydrogenlike ion. The ion with a spinless nucleus is assumed to be placed into the homogeneous magnetic field $\boldsymbol{\mathcal{H}}$ described by the classical vector potential of the form $\bfA_\text{cl} (\bfr) = [\boldsymbol{\mathcal{H}} \times \bfr ] /2$. Within the zeroth-order approximation, the electron obeys the Dirac equation with the spherically symmetric binding potential of the pointlike nucleus $V(r)=-\alpha Z/r$,
\begin{align}
\label{eq:Dirac}
h^{\rm D} |n\rangle \equiv
\big( 
\balpha\cdot\bfp + \beta m + V
\big) 
|n\rangle
= \veps_n 
|n\rangle  \, ,
\end{align}
where $\balpha$ and $\beta$ are the Dirac matrices and $\bfp$ is the momentum operator. By replacing $V$ in Eq.~(\ref{eq:Dirac}) with the potential of the extended nucleus, one can partially take into account the nuclear size correction to the recoil effect. For simplicity, we direct the $z$ axis along the magnetic field, $\boldsymbol{\mathcal{H}}=\mathcal{H}{\bf e}_z$. Then, the contribution to the Dirac Hamiltonian due to the coupling with $\boldsymbol{\mathcal{H}}$ reads as follows: $-e\balpha\cdot\bfA_\text{cl}(\bfr) = \mu_0\mathcal{H}{m}\,[ \bfr \times \balpha ]_z$, where $\mu_0=|e|/2m$ is the Bohr magneton. 
According to Ref.~\cite{Shabaev:2001:052104}, the nuclear recoil contribution to the $g$ factor of the state~$|a\rangle$ with the Dirac energy $\veps_a$ and the angular momentum projection $m_a$ is conveniently represented by the sum of two terms, $\Delta g = \Delta g_{\rm L} + \Delta g_{\rm H}$, where  
\begin{widetext}
\begin{align}
\label{eq:g_L}
\Delta g_{\rm L} =&\, \frac{1}{m_a} \frac{{m}}{M} \,
  \left\{
      \la \delta a| \Big( \bfp^2 - 2 \bfp \cdot \bfD(0) \Big) | a \ra 
    - \la a | \Big( [ \bfr\times\bfp ]_z - [ \bfr\times\bfD(0) ]_z \Big) | a \ra 
  \right\}  \, ,   \\
\label{eq:g_H}
\Delta g_{\rm H} =&\, \frac{1}{m_a} \frac{m}{M} \frac{i}{2\pi}
    \int_{-\infty}^{\infty} \! d\omega \,
    \Bigl\{ \la \delta a | B^k_{-}(\omega) G(\omega+\veps_a) B^k_{+}(\omega) | a \ra          
                 + \la a | B^k_{-}(\omega) G(\omega+\veps_a) B^k_{+}(\omega) | \delta a \ra   \nonumber\\ 
& \qquad\qquad\qquad\quad\,\,
+ \la a | B^k_{-}(\omega) G(\omega+\veps_a) 
          \Big( [ \bfr\times\balpha ]_z - \langle a | [\bfr\times\balpha]_z | a \rangle \Big) 
                          G(\omega+\veps_a) B^k_{+}(\omega) | a \ra \Bigr\} \, .  
\end{align}
\end{widetext}
Here $|\delta a\ra = \sum_n^{\veps_n\ne \veps_a} | n \ra \la n | [\bfr\times\balpha]_z | a \ra (\veps_a-\veps_n)^{-1}$ is the wave-function correction due to the external magnetic field, $G(\omega)=\sum_n|n\ra \la n|[\omega-\veps_n(1-i0)]^{-1}$ is the Dirac-Coulomb Green's function, $B^k_{\pm}(\omega)=D^k(\omega)\pm [p^k,V]/(\omega+i0)$, $[A,B]=AB-BA$, $D^k(\omega)=-4\pi\alpha Z\alpha^l D^{lk}(\omega)$, and
\begin{align}
\label{eq:D_lk}
D^{lk}(\omega,\bfr) =&\, 
-\frac{1}{4\pi} \left[ 
\frac{\exp\left( i \sqrt{\omega^2 + i0} \, r \right)}{ r } \delta_{lk} \right. \nonumber\\
& \qquad\,
+  \left.
\nabla^l \nabla^k
\frac{\exp\left( i \sqrt{\omega^2 + i0} \, r \right) - 1}{\omega^2 r}
\right] 
\end{align}
is the transverse part of the photon propagator in the Coulomb gauge with the branch of the square root fixed by the condition $\Im\left(\sqrt{\omega^2 + i0}\right)>0$. The summation over the repeated indices is implied. The zero-energy-transfer limit $\omega\to 0$ of the vector $D^k(\omega)$ appearing in Eq.~(\ref{eq:g_L}) has the form
\begin{align}
\label{eq:D0}
\bfD(0) = \frac{\alpha Z}{2r} \left[ \balpha + \frac{(\balpha\cdot\bfr)}{r^2}\,\bfr \right] \, .
\end{align}
Therefore, the vector product $[ \bfr\times\bfD(0) ]_z$ in Eq.~(\ref{eq:g_L}) can be also written as $\alpha Z [ \bfr\times\balpha ]_z /2r$.

The low-order contribution $\Delta g_{\rm L}$ can be derived from the relativistic Breit equation. The operators $\bfp^2$ and $[ \bfr\times\bfp ]_z\equiv l_z$ ($l_z$ is the orbital angular momentum) in Eq.~(\ref{eq:g_L}) correspond to the nonrelativistic limit whereas the terms with the vector $\bfD(0)$ provide the lowest-order relativistic correction. In the meantime, the derivation of the higher-order part $\Delta g_{\rm H}$ requires application of bound-state QED beyond the Breit approximation. For this reason, in the following we will refer to this part as the QED one. We should note that the formalism developed in Ref.~\cite{Shabaev:2001:052104} can be easily adopted to treat the nuclear recoil effect on the bound-electron $g$ factors of ions with one electron over the closed shells. To this end, the representation in which the closed shells are regarded as belonging to the vacuum is to be employed, see, e.g., Refs.~\cite{Shabaev:2002:062104, Shabaev:2017:263001}.


\begin{figure}[b!]
\begin{center}
\includegraphics[width=0.8\linewidth]{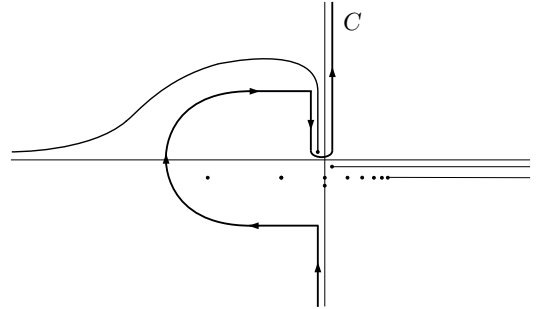}
\caption{\label{fig:contour}
The poles and the branch cuts of the integrand for the part with $| \delta a \rangle$ of the one-transverse-photon contribution, and the integration contour~$C$ used for the evaluation of this correction. 
}
\end{center}
\vspace*{-4mm}
\end{figure}


In the case of the pointlike nucleus which is considered in the present study, the calculations of the low-order part $\Delta g_{\rm L}$ can be performed analytically for arbitrary state of the hydrogenlike ion. The operators $\bfp^2$ and $\bfp \cdot \bfD(0)$ in Eq.~(\ref{eq:g_L}) are invariant under rotation. Therefore, only the component of $|\delta a\rangle$ possessing the same angular quantum numbers as the unperturbed wave function $|a\rangle$ contributes. This component can be obtained by employing the generalized virial relations for the Dirac equation~\cite{Shabaev:1991:4479}, which result in
\begin{align}
\label{eq:wf_da}
| \delta a \rangle_{\kappa m_a}
=
\dvec{ X(r) \Omega_{\kappa m_a} (\bfnr) }{ i Y(r) \Omega_{-\kappa m_a} (\bfnr) }  \, ,
\end{align}
where
\begin{align}
\label{eq:wf_X}
X(r) =&\, 
   \frac{\kappa m_a}{j(j+1)} \left\{
   \left[ \frac{2\kappa(m+\veps_a)-m}{2m^2} \, r + \frac{\kappa\alpha Z}{m^2} \right] f(r) \right. \nonumber \\
& \qquad\qquad
        \left. + \frac{\kappa-2\kappa^2}{2m^2} \, g(r) \right\} \, ,   \\
\label{eq:wf_Y}   
Y(r) =&\, 
   \frac{\kappa m_a}{j(j+1)} \left\{
   \left[ \frac{2\kappa(m-\veps_a)+m}{2m^2} \, r - \frac{\kappa\alpha Z}{m^2} \right] g(r) \right. \nonumber \\
& \qquad\qquad
        \left. + \frac{\kappa+2\kappa^2}{2m^2} \, f(r) \right\} \, .  
\end{align}
Here $\kappa$ is the Dirac angular quantum number of the state~$|a\rangle$, $j=|\kappa|-1/2$ is the total angular momentum, and $g$ and $f$ are the large and small radial components of the unperturbed wave function 
\begin{align}
\label{eq:wf_a}
| a \rangle =
\dvec{ g(r) \Omega_{\kappa m_a} (\bfnr) }{ i f(r) \Omega_{-\kappa m_a} (\bfnr) } \, .
\end{align}
Applying the formulas presented in Ref.~\cite{Shabaev:1991:4479}, one can obtain the following expression for the low-order part of the nuclear recoil contribution to the bound-electron $g$ factor~\cite{Shabaev:2001:052104} in the point-nucleus case,
\begin{align}
\label{eq:g_L_point}
\Delta g_{\rm L} = -\frac{m}{M} \, \frac{2\kappa^2 \varepsilon^2_a + \kappa m \varepsilon_a - m^2}{2m^2 j (j+1)} \, .
\end{align}
For the $n=1$ and $n=2$ states, Eq.~(\ref{eq:g_L_point}) leads to
\begin{align}
\label{eq:g_L_1s}\!\!\!
\Delta g_{\rm L}^{1s} &= 
\frac{m}{M}\,\frac{2}{3} (1-\gamma_1)(1+2\gamma_1) \, , \\
\label{eq:g_L_2s}\!\!\!
\Delta g_{\rm L}^{2s} &= 
\frac{m}{M}\,\frac{1}{3} \big(2-\sqrt{2(1+\gamma_1)}\big)\big(1+\sqrt{2(1+\gamma_1)}\big)  , \\
\label{eq:g_L_2p1}\!\!\!
\Delta g_{\rm L}^{2p_{1/2}} &= 
\frac{m}{M}\,\frac{1}{3} \big(2+\sqrt{2(1+\gamma_1)}\big)\big(1-\sqrt{2(1+\gamma_1)}\big)  , \\
\label{eq:g_L_2p3}\!\!\!
\Delta g_{\rm L}^{2p_{3/2}} &= 
\frac{m}{M}\,\frac{2}{15} (1-\gamma_2)(1+2\gamma_2) \, ,
\end{align}
where $\gamma_1=\sqrt{1-(\alpha Z)^2}$ and $\gamma_2=\sqrt{4-(\alpha Z)^2}$. The leading orders in $\alpha Z$ read as
\begin{align}
\label{eq:g_L_1s_expan}
\Delta g_{\rm L}^{1s} &= 
\frac{m}{M} \left[ (\alpha Z)^2 - \frac{1}{12}(\alpha Z)^4 + \ldots \right] \, , \\
\label{eq:g_L_2s_expan}
\Delta g_{\rm L}^{2s} &= 
\frac{m}{M} \left[ \frac{1}{4}(\alpha Z)^2 + \frac{11}{192}(\alpha Z)^4 + \ldots \right] \, , \\
\label{eq:g_L_2p1_expan}
\Delta g_{\rm L}^{2p_{1/2}} &= 
\frac{m}{M} \left[ -\frac{4}{3} +\frac{5}{12}(\alpha Z)^2 + \ldots \right] \, , \\
\label{eq:g_L_2p3_expan}
\Delta g_{\rm L}^{2p_{3/2}} &= 
\frac{m}{M} \left[ -\frac{2}{3} +\frac{7}{30}(\alpha Z)^2 + \ldots \right] \, .
\end{align}
It is seen that for the $s$ states $(\kappa=-1)$ the nonrelativistic contribution to $\Delta g_{\rm L}$ vanishes, and the $\alpha Z$-expansion starts with the term of the pure relativistic $[\sim (\alpha Z)^2]$ origin. For the $p$ states ($\kappa=1$ or $\kappa=-2$), there is a nonzero nonrelativistic limit of the nuclear recoil effect on the bound-electron $g$ factor.

The higher-order part $\Delta g_{\rm H}$ is evaluated numerically. It is naturally divided into three contributions depending on the number of the $\bfD$ vectors. The term without $\bfD$ is referred to as the ``Coulomb'' ($\rm Coul$) contribution while the terms including one and two $\bfD$ vectors are termed as the ``one-transverse-photon'' ($\rm tr1$) and ``two-transverse-photon'' ($\rm tr2$) contributions, respectively. The $\omega$ integration for the simplest Coulomb contribution can be carried out analytically by employing Cauchy's residue theorem,
\begin{widetext}
\begin{align}
\label{eq:g_H_coul}
\Delta g_{\rm H}^{\rm Coul} =&\,
\frac{1}{m_a}\frac{m}{M} \Bigg\{ 
\sum_{n<0}
\frac{ \langle \delta a | [p^k,V] | n \rangle \langle n | [p^k,V] | a \rangle 
     + \langle a | [p^k,V] | n \rangle \langle n | [p^k,V] | \delta a \rangle }{(\veps_a-\veps_n)^2}   \nonumber \\
&\qquad 
+ 2 \sum_{n<0}
\frac{ \langle a | [p^k,V] | n \rangle 
       \langle n | \Big( [ \bfr\times\balpha ]_z - \langle a | [\bfr\times\balpha]_z | a \rangle \Big) | n \rangle
       \langle n | [p^k,V] | a \rangle }{(\veps_a-\veps_n)^3}     \nonumber \\
&\qquad 
+ \sum_{n_1<0} \!\! \sum_{n_2}^{\veps_{n_2}\neq\veps_{n_1}}
\frac{ \langle a   | [p^k,V] | n_1 \rangle
       \langle n_1 | [ \bfr\times\balpha ]_z | n_2 \rangle
       \langle n_2 | [p^k,V] | a   \rangle }{(\veps_a-\veps_{n_1})^2(\veps_{n_1}-\veps_{n_2})}   \nonumber \\ 
  &\qquad 
+ \sum_{n_2<0} \!\! \sum_{n_1}^{\veps_{n_1}\neq\veps_{n_2}}
\frac{ \langle a   | [p^k,V] | n_1 \rangle
       \langle n_1 | [ \bfr\times\balpha ]_z | n_2 \rangle
       \langle n_2 | [p^k,V] | a   \rangle }{(\veps_a-\veps_{n_2})^2(\veps_{n_2}-\veps_{n_1})} 
\Bigg\} \, ,
\end{align}
\end{widetext}
where the notation $n<0$ implies that the corresponding summation runs over the negative-energy part of the spectrum only, $\veps_{n}\leqslant-mc^2$. The $\omega$~integration for the $\Delta g_{\rm H}^{\rm tr1}$ and $\Delta g_{\rm H}^{\rm tr2}$ terms is performed numerically using Wick's rotation. An example of the integration contour employed in the present calculations is shown in Fig.~\ref{fig:contour}. The branch cuts of the photon propagator (\ref{eq:D_lk}), the poles of the Green's function $G(\omega+\veps_a)$, and the pole $1/(\omega+i0)$ of the vector $B^k(\omega)$ are depicted as well. The contour is chosen to avoid the singularities near $\omega=0$ and go around the poles of the bound states with $\veps_n<\veps_a$. This is done since particular care is required at low values of the integration variable~$\omega$. As it is for the low-order part $\Delta g_{\rm L}$, the expression sandwiched between $|a\rangle$ and $|\delta a\rangle$ in Eq.~(\ref{eq:g_H}) conserves the angular quantum numbers. For this reason, the Eqs.~(\ref{eq:wf_da})--(\ref{eq:wf_Y}) can be also employed to calculate the corresponding contribution to the higher-order part. Finally, the summation over the intermediate electron states is carried out using the finite basis sets constructed from $\rm B$~splines~\cite{Johnson:1988:307, Sapirstein:1996:5213}.

\section{Results and discussion}


\begin{table*}[h!]
\begin{minipage}{0.485\textwidth}
  \centering

{
\renewcommand{\arraystretch}{1.20}

\caption{\label{tab:g_rec_1s}
The higher-order (QED) nuclear recoil contribution to the $g$ factor of the $1s$ state.
The results are expressed in terms of the function $P^{(5|1)}(\alpha Z)$ defined by Eq.~(\ref{eq:P}).
The individual terms of $P^{(5|1)}(\alpha Z)=P^{(5|1)}_{\rm Coul}(\alpha Z)+P^{(5|1)}_{\rm tr1}(\alpha Z)+P^{(5|1)}_{\rm tr2}(\alpha Z)$ are shown.
}

\begin{tabular}{r@{\quad}
                S[table-format=-2.6]
                S[table-format= 4.6]
                S[table-format=-3.6]
                S[table-format= 3.6]
                @{}
               }
               
\hline
\hline \\[-3.5mm]

  $Z$  &  {$P^{(5|1)}_{\rm Coul}(\alpha Z)$}
       &  {$P^{(5|1)}_{\rm tr1 }(\alpha Z)$}
       &  {$P^{(5|1)}_{\rm tr2 }(\alpha Z)$}
       &  {$P^{(5|1)}_{\rm     }(\alpha Z)$}  \\
        
\hline \\[-4mm]

  1  &      -1.11414  &     100.70120  &   -80.82002  &      18.76704    \\ 
  2  &      -1.09754  &      53.52779  &   -36.98689  &      15.44337    \\ 
  3  &      -1.08183  &      37.44950  &   -22.80837  &      13.55930    \\ 
  4  &      -1.06693  &      29.24593  &   -15.91960  &      12.25940    \\ 
  5  &      -1.05277  &      24.23028  &   -11.90049  &      11.27702    \\ 
  6  &      -1.03931  &      20.82713  &    -9.29387  &      10.49396    \\ 
  7  &      -1.02649  &      18.35587  &    -7.48193  &       9.84744    \\ 
  8  &      -1.01429  &      16.47349  &    -6.15902  &       9.30018    \\ 
  9  &      -1.00267  &      14.98800  &    -5.15711  &       8.82821    \\ 
 10  &      -0.99161  &      13.78331  &    -4.37646  &       8.41524    \\ 
 11  &      -0.98106  &      12.78501  &    -3.75425  &       8.04970    \\ 
 12  &      -0.97102  &      11.94311  &    -3.24902  &       7.72307    \\ 
 13  &      -0.96145  &      11.22277  &    -2.83240  &       7.42892    \\ 
 14  &      -0.95235  &      10.59892  &    -2.48431  &       7.16226    \\ 
 15  &      -0.94368  &      10.05304  &    -2.19020  &       6.91916    \\ 
 16  &      -0.93544  &       9.57116  &    -1.93926  &       6.69645    \\ 
 17  &      -0.92762  &       9.14249  &    -1.72332  &       6.49156    \\ 
 18  &      -0.92019  &       8.75863  &    -1.53608  &       6.30236    \\ 
 19  &      -0.91314  &       8.41286  &    -1.37263  &       6.12709    \\ 
 20  &      -0.90647  &       8.09979  &    -1.22907  &       5.96425    \\ 

\hline
\hline

\end{tabular}%

}


\end{minipage}
\hfill
\begin{minipage}{0.485\textwidth}
  \centering

{
\renewcommand{\arraystretch}{1.20}

\caption{\label{tab:g_rec_2s}
The higher-order (QED) nuclear recoil contribution to the $g$ factor of the $2s$ state.
The results are expressed in terms of the function $P^{(5|2)}(\alpha Z)$ defined by Eq.~(\ref{eq:P}).
The individual terms of $P^{(5|2)}(\alpha Z)=P^{(5|2)}_{\rm Coul}(\alpha Z)+P^{(5|2)}_{\rm tr1}(\alpha Z)+P^{(5|2)}_{\rm tr2}(\alpha Z)$ are shown.
}

\begin{tabular}{r@{\quad}
                S[table-format=-2.6]
                S[table-format= 4.5(1)]
                S[table-format=-3.6]
                S[table-format= 3.5(1)]
                @{}
               }
               
\hline
\hline \\[-3.5mm]

  $Z$  &  {$P^{(5|2)}_{\rm Coul}(\alpha Z)$}
       &  {$P^{(5|2)}_{\rm tr1 }(\alpha Z)$}
       &  {$P^{(5|2)}_{\rm tr2 }(\alpha Z)$}
       &  {$P^{(5|2)}_{\rm     }(\alpha Z)$}  \\
        
\hline \\[-4mm]

  1  &      -1.11417  &  100.96878(1)  &   -80.65709  &   19.19753(1)    \\ 
  2  &      -1.09764  &      53.79672  &   -36.82355  &      15.87553    \\ 
  3  &      -1.08207  &      37.72011  &   -22.64433  &      13.99371    \\ 
  4  &      -1.06736  &      29.51850  &   -15.75462  &      12.69653    \\ 
  5  &      -1.05344  &      24.50508  &   -11.73436  &      11.71728    \\ 
  6  &      -1.04028  &      21.10440  &    -9.12641  &      10.93772    \\ 
  7  &      -1.02781  &      18.63583  &    -7.31296  &      10.29505    \\ 
  8  &      -1.01602  &      16.75636  &    -5.98840  &       9.75195    \\ 
  9  &      -1.00485  &      15.27398  &    -4.98470  &       9.28443    \\ 
 10  &      -0.99428  &      14.07259  &    -4.20212  &       8.87619    \\ 
 11  &      -0.98429  &      13.07778  &    -3.57784  &       8.51565    \\ 
 12  &      -0.97485  &      12.23956  &    -3.07044  &       8.19427    \\ 
 13  &      -0.96594  &      11.52308  &    -2.65152  &       7.90562    \\ 
 14  &      -0.95754  &      10.90328  &    -2.30103  &       7.64471    \\ 
 15  &      -0.94963  &      10.36162  &    -2.00441  &       7.40758    \\ 
 16  &      -0.94219  &       9.88413  &    -1.75086  &       7.19109    \\ 
 17  &      -0.93522  &       9.46005  &    -1.53220  &       6.99264    \\ 
 18  &      -0.92869  &       9.08095  &    -1.34214  &       6.81012    \\ 
 19  &      -0.92260  &       8.74012  &    -1.17577  &       6.64174    \\ 
 20  &      -0.91693  &       8.43217  &    -1.02921  &       6.48603    \\ 

\hline
\hline

\end{tabular}%

}


\end{minipage}
\end{table*}

\begin{figure*}[h!]
\begin{minipage}[b]{.485\textwidth}
  \includegraphics[width=\linewidth]{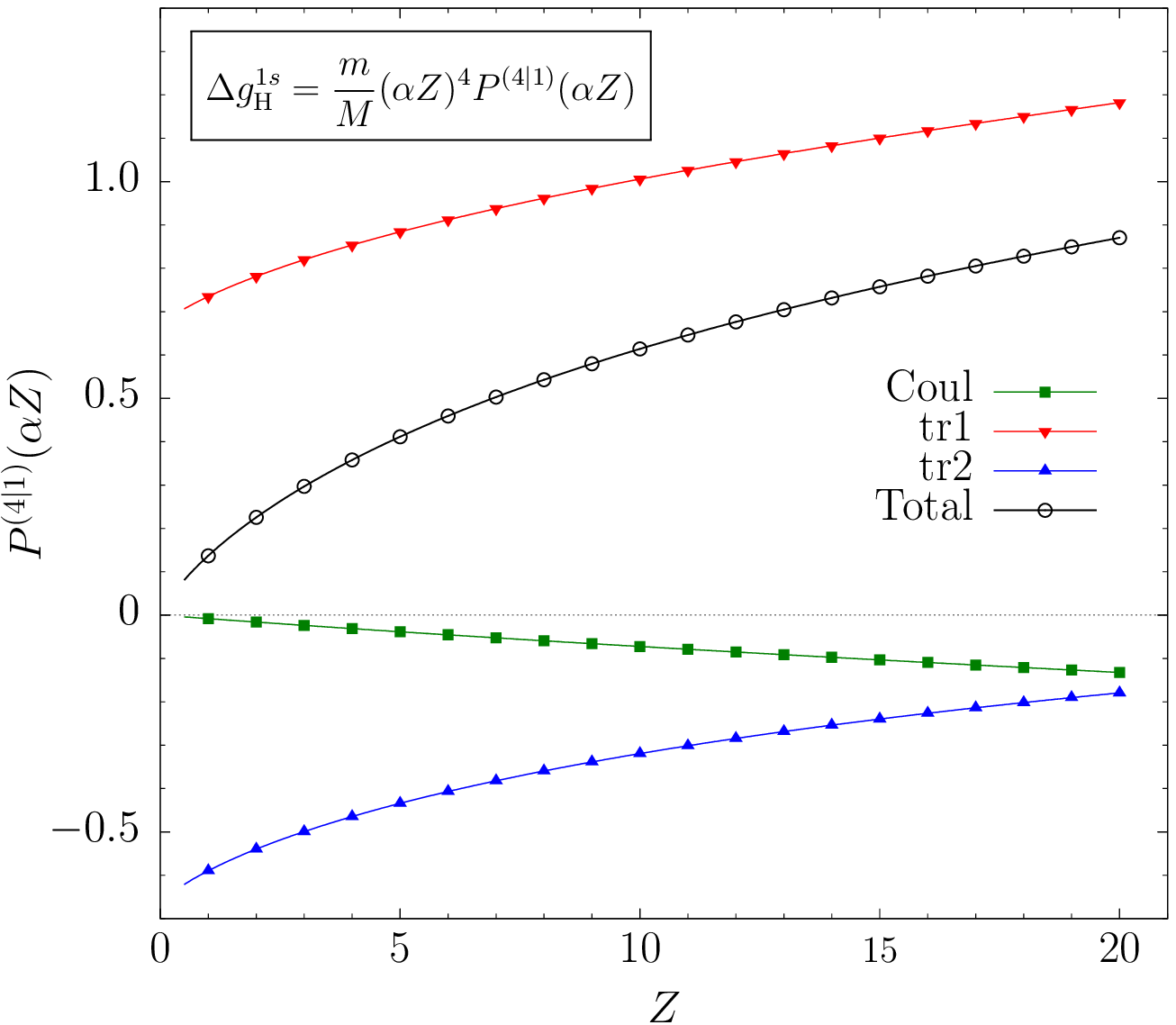}
  \caption{\label{fig:1s} The Coulomb, one-transverse-photon, and two-transverse-photon contributions to the higher-order nuclear recoil effect on the $1s$ $g$ factor. The results are presented in terms of the function $P^{(4|1)}(\alpha Z)$ defined by Eq.~(\ref{eq:P}). 
  Note that $P^{(4|1)}(x) = x P^{(5|1)}(x)$.}
\end{minipage}
\hfill
\begin{minipage}[b]{.485\textwidth}
  \includegraphics[width=\linewidth]{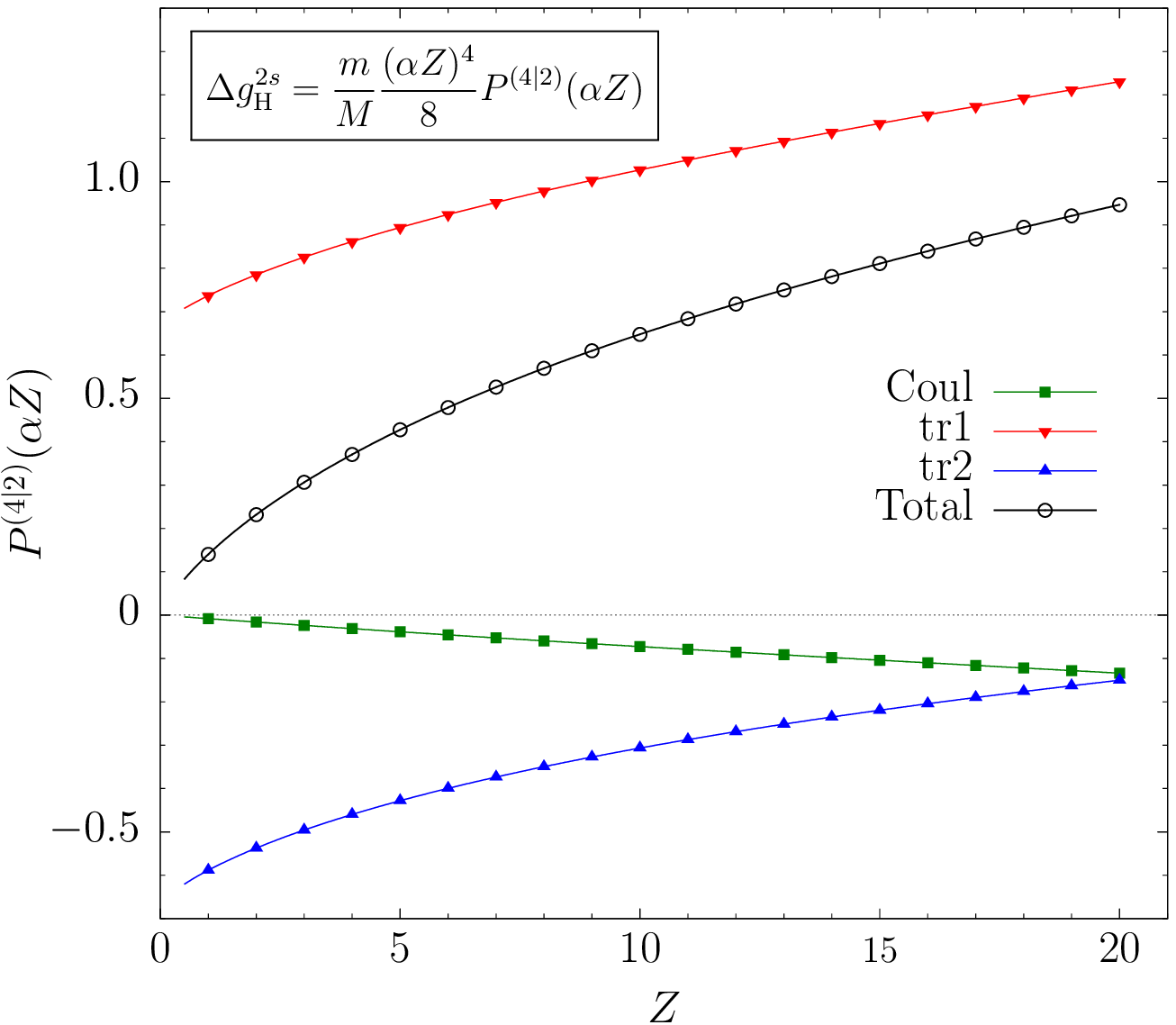}
  \caption{\label{fig:2s} The Coulomb, one-transverse-photon, and two-transverse-photon contributions to the higher-order nuclear recoil effect on the $2s$ $g$ factor. The results are presented in terms of the function $P^{(4|2)}(\alpha Z)$ defined by Eq.~(\ref{eq:P}).
  Note that $P^{(4|2)}(x) = x P^{(5|2)}(x)$.}
\end{minipage}
\end{figure*}


In this section, we present our results for the nontrivial QED part of the nuclear recoil effect on the bound-electron $g$ factor of the $1s$, $2s$, $2p_{1/2}$, and $2p_{3/2}$ states in hydrogenlike ions with $Z=1-20$ evaluated for pointlike nuclei. For further consideration, it is useful to introduce the dimensionless functions $P^{(k|n)}(\alpha Z)$ defined as
\begin{align}
\label{eq:P}
\Delta g_{\rm H} = \frac{m}{M} \frac{(\alpha Z)^k}{n^3} P^{(k|n)}(\alpha Z) \, ,
\end{align}
where $n$ is the principal quantum number and arbitrary integer $k$ can be chosen for the convenient representation of the results.

The higher-order nuclear recoil contributions to the $1s$ and $2s$ $g$ factors are presented in Tables~\ref{tab:g_rec_1s} and \ref{tab:g_rec_2s}, respectively. The results are shown in terms of the function $P^{(5|1)}(\alpha Z)$ for the $1s$ state and $P^{(5|2)}(\alpha Z)$ for the $2s$ state. For particular ions, this contribution was considered earlier in Refs.~\cite{Shabaev:2002:091801, Kohler:2016:10246, Shabaev:2017:263001}. 
Our present results are in agreement with the previous ones but are given to a higher accuracy. The uncertainties are estimated by studying the convergence of the $\omega$ integration in Eq.~(\ref{eq:g_H}) as well as by increasing the size of the basis employed. When the uncertainty is not specified, all the digits presented are assumed to be correct. 

In Ref.~\cite{Shabaev:2002:091801}, the behavior of the higher-order contribution $\Delta g_{\rm H}$ for the $1s$ state as a function of $\alpha Z$ when $Z$ tends to zero was studied. It was found that the total result exhibits the $(\alpha Z)^5$ behavior, whereas the one-transverse-photon and two-transverse-photon terms taken separately behave as $(\alpha Z)^4$. Moreover, the individual contributions to $\Delta g_{\rm H}^{\rm tr1}$, namely, the part with and without $|\delta a \rangle$, include even the lower power of $\alpha Z$ and manifest the $(\alpha Z)^3$ behavior. In the present work, we study the QED recoil contribution to the $1s$ and $2s$ $g$ factors for small $Z$. It turns out that the higher-order part of the nuclear recoil effect~$\Delta g_{\rm H}$ is rather similar for the $g$ factors of both $s$ states. This fact is clearly demonstrated in Figs.~\ref{fig:1s} and \ref{fig:2s} where the Coulomb, one-transverse-photon, and two-transverse-photon contributions as well as the total values of the $\Delta g_{\rm H}$ correction are plotted for the $1s$ and $2s$ states in terms of the functions $P^{(4|1)}(\alpha Z)$ and $P^{(4|2)}(\alpha Z)$, respectively. One can see that for both states these functions for the $\Delta g_{\rm H}^{\rm tr1}$ and $\Delta g_{\rm H}^{\rm tr2}$ terms possess nonzero limits at $\alpha Z \rightarrow 0$ which cancel each other in the sum. The appearance of the curves is almost the same. We have performed our calculations for a series of $Z$ including fractional values and fitted the results using the least-squares method to the form
\begin{align}
\label{eq:fit_1s}
P^{(5|1)}_{1s}(\alpha Z) &= A_{1s}^{51} \log(\alpha Z) + A_{1s}^{50} + \alpha Z (\ldots) \, ,   \\
\label{eq:fit_2s}
P^{(5|2)}_{2s}(\alpha Z) &= A_{2s}^{51} \log(\alpha Z) + A_{2s}^{50} + \alpha Z (\ldots) \, .
\end{align}
By analyzing the dependence of the results on the number of the varying parameters in the fit and the number of the fitting points, we have found that for the $1s$ state $A_{1s}^{51}=-5.1(2)$ and $A_{1s}^{50}=-6.6(5)$ and for the $2s$ state $A_{2s}^{51}=-5.1(2)$ and $A_{2s}^{50}=-6.2(5)$. The coefficients obtained for the $1s$ state are in agreement with those of Ref.~\cite{Shabaev:2002:091801} but have higher accuracy. 


\begin{figure}[h!]
  \includegraphics[width=\linewidth]{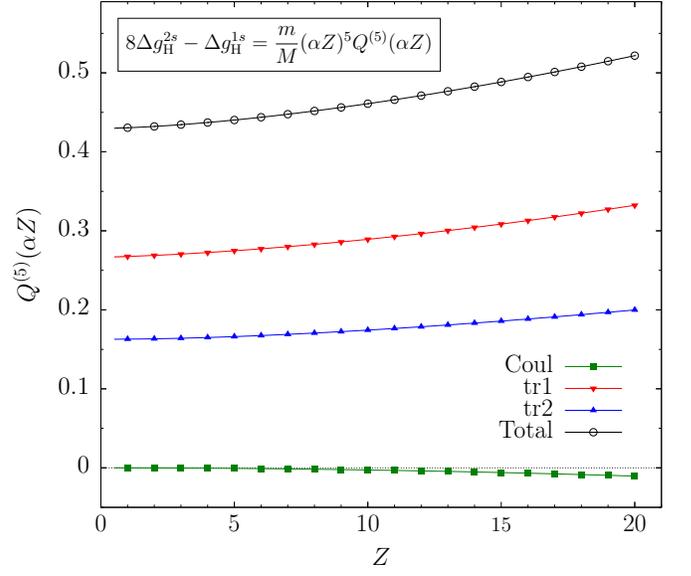}
  \caption{\label{fig:2s-1s} The Coulomb, one-transverse-photon, and two-transverse-photon terms of the weighted difference of the higher-order nuclear recoil contributions to the $g$ factors of the $2s$ and $1s$ states, $8\Delta g_{\rm H}^{2s}-\Delta g_{\rm H}^{1s}$. The results are presented in terms of the function $Q^{(5)}(\alpha Z)$ defined by Eq.~(\ref{eq:Q}).}
\end{figure}


Since the coefficients of the logarithmic terms for the $1s$ and $2s$ states in Eqs.~(\ref{eq:fit_1s}) and (\ref{eq:fit_2s}) are the same, at least within the numerical uncertainty of the present fit, it is also useful to consider the weighted difference $\eta \equiv 8\Delta g_{\rm H}^{2s}-\Delta g_{\rm H}^{1s}$ [we remind that, compared to the $1s$ state, for the $2s$ state the additional factor $1/8$ is separated in the definition of the function $P(\alpha Z)$]. In Fig.~\ref{fig:2s-1s}, the difference $\eta$ is plotted together with the individual contributions to it in terms of the function $Q^{(5)}(\alpha Z)$ defined according to
\begin{align}
\label{eq:Q}
\eta &=
\frac{m}{M} (\alpha Z)^5 Q^{(5)}(\alpha Z) \, ,  \\
\label{eq:Q_P}
Q^{(5)}(\alpha Z) &= 
P^{(5|2)}_{2s}(\alpha Z) - P^{(5|1)}_{1s}(\alpha Z) \, .
\end{align}
The plots in Fig.~\ref{fig:2s-1s} clearly show that the logarithmic terms indeed cancel each other in this difference. Moreover, the terms of the order $(\alpha Z)^4$ vanish in the one-transverse-photon and two-transverse-photon contributions into the difference $\eta$. Finally, the leading terms of the order $(\alpha Z)^5$ in the Coulomb parts of $\Delta g_{\rm H}^{1s}$ and $\Delta g_{\rm H}^{2s}$ also cancel each other. Therefore, the limit of $Q^{(5)}(\alpha Z)$ at $\alpha Z\rightarrow 0$ is finite and it is related with the coefficients $A_{1s}^{50}$ and $A_{2s}^{50}$ in Eqs.~(\ref{eq:fit_1s}) and (\ref{eq:fit_2s}) as follows
\begin{align}
Q^{(5)}(0) = A_{2s}^{50} - A_{1s}^{50} \, .
\end{align}
The limit of the function $Q^{(5)}(\alpha Z)$ at $\alpha Z \rightarrow 0$ can be determined by the least-squares fitting. We obtain $Q^{(5)}(0)=0.43$ for the total value of the weighted difference~$\eta$ and $Q^{(5)}_{\rm Coul}(0)\equiv 0$, $Q^{(5)}_{\rm tr1}(0)=0.27$, $Q^{(5)}_{\rm tr2}(0)=0.16$ for the Coulomb, one-transverse-photon, and two-transverse-photon contributions, respectively.


\begin{table*}[h!]
\begin{minipage}{0.485\textwidth}
  \centering

{
\renewcommand{\arraystretch}{1.20}

\caption{\label{tab:g_rec_2p1}
The higher-order (QED) nuclear recoil contribution to the $g$ factor of the $2p_{1/2}$ state.
The results are presented in terms of the function $P^{(3|2)}(\alpha Z)$ defined by Eq.~(\ref{eq:P}).
The individual terms of $P^{(3|2)}(\alpha Z)=P^{(3|2)}_{\rm Coul}(\alpha Z)+P^{(3|2)}_{\rm tr1}(\alpha Z)+P^{(3|2)}_{\rm tr2}(\alpha Z)$ are shown.
}

\begin{tabular}{r@{\quad}
                S[table-format=-2.8]
                S[table-format= 2.7]
                S[table-format= 2.7]
                S[table-format= 2.7]
                @{}
               }
               
\hline
\hline \\[-3.5mm]

  $Z$  &  {$P^{(3|2)}_{\rm Coul}(\alpha Z)$}
       &  {$P^{(3|2)}_{\rm tr1 }(\alpha Z)$}
       &  {$P^{(3|2)}_{\rm tr2 }(\alpha Z)$}
       &  {$P^{(3|2)}_{\rm     }(\alpha Z)$}  \\
        
\hline \\[-4mm]

  1  &  \multicolumn{1}{l}{\num[{scientific-notation=true,round-mode=places,round-precision=2}]{-0.000000001778}}  &    0.421036  &    0.003339  &    0.424375    \\ 
  2  &  \multicolumn{1}{l}{\num[{scientific-notation=true,round-mode=places,round-precision=2}]{-0.000000027602}}  &    0.424393  &    0.006814  &    0.431206    \\ 
  3  &  \multicolumn{1}{l}{\num[{scientific-notation=true,round-mode=places,round-precision=2}]{-0.000000136082}}  &    0.427798  &    0.010393  &    0.438191    \\ 
  4  &  \multicolumn{1}{l}{\num[{scientific-notation=true,round-mode=places,round-precision=2}]{-0.000000419872}}  &    0.431243  &    0.014061  &    0.445303    \\ 
  5  &  \multicolumn{1}{l}{\num[{scientific-notation=true,round-mode=places,round-precision=2}]{-0.000001002612}}  &    0.434721  &    0.017806  &    0.452526    \\ 
  6  &  \multicolumn{1}{l}{\num[{scientific-notation=true,round-mode=places,round-precision=2}]{-0.000002036556}}  &    0.438227  &    0.021619  &    0.459844    \\ 
  7  &  \multicolumn{1}{l}{\num[{scientific-notation=true,round-mode=places,round-precision=2}]{-0.000003700721}}  &    0.441759  &    0.025493  &    0.467249    \\ 
  8  &  \multicolumn{1}{l}{\num[{scientific-notation=true,round-mode=places,round-precision=2}]{-0.000006199432}}  &    0.445314  &    0.029422  &    0.474730    \\ 
  9  &  \multicolumn{1}{l}{\num[{scientific-notation=true,round-mode=places,round-precision=2}]{-0.000009761194}}  &    0.448889  &    0.033401  &    0.482280    \\ 
 10  &  \multicolumn{1}{l}{\num[{scientific-notation=true,round-mode=places,round-precision=2}]{-0.000014637852}}  &    0.452483  &    0.037425  &    0.489894    \\ 
 11  &  \multicolumn{1}{l}{\num[{scientific-notation=true,round-mode=places,round-precision=2}]{-0.000021103990}}  &    0.456094  &    0.041491  &    0.497564    \\ 
 12  &  \multicolumn{1}{l}{\num[{scientific-notation=true,round-mode=places,round-precision=2}]{-0.000029456543}}  &    0.459722  &    0.045595  &    0.505287    \\ 
 13  &  \multicolumn{1}{l}{\num[{scientific-notation=true,round-mode=places,round-precision=2}]{-0.000040014607}}  &    0.463364  &    0.049734  &    0.513057    \\ 
 14  &  \multicolumn{1}{l}{\num[{scientific-notation=true,round-mode=places,round-precision=2}]{-0.000053119423}}  &    0.467021  &    0.053904  &    0.520872    \\ 
 15  &  \multicolumn{1}{l}{\num[{scientific-notation=true,round-mode=places,round-precision=2}]{-0.000069134532}}  &    0.470692  &    0.058104  &    0.528727    \\ 
 16  &  \multicolumn{1}{l}{\num[{scientific-notation=true,round-mode=places,round-precision=2}]{-0.000088446079}}  &    0.474376  &    0.062331  &    0.536619    \\ 
 17  &   -0.000111  &    0.478074  &    0.066583  &    0.544546    \\ 
 18  &   -0.000139  &    0.481786  &    0.070857  &    0.552505    \\ 
 19  &   -0.000170  &    0.485511  &    0.075153  &    0.560494    \\ 
 20  &   -0.000207  &    0.489251  &    0.079468  &    0.568511    \\ 

\hline
\hline

\end{tabular}%

}


\end{minipage}
\hfill
\begin{minipage}{0.485\textwidth}
  \centering

{
\renewcommand{\arraystretch}{1.20}

\caption{\label{tab:g_rec_2p3}
The higher-order (QED) nuclear recoil contribution to the $g$ factor of the $2p_{3/2}$ state.
The results are presented in terms of the function $P^{(3|2)}(\alpha Z)$ defined by Eq.~(\ref{eq:P}).
The individual terms of $P^{(3|2)}(\alpha Z)=P^{(3|2)}_{\rm Coul}(\alpha Z)+P^{(3|2)}_{\rm tr1}(\alpha Z)+P^{(3|2)}_{\rm tr2}(\alpha Z)$ are shown.
}

\begin{tabular}{r@{\quad}
                S[table-format=-2.8]
                S[table-format= 2.7]
                S[table-format= 2.7]
                S[table-format= 2.7]
                @{}
               }
               
\hline
\hline \\[-3.5mm]

  $Z$  &  {$P^{(3|2)}_{\rm Coul}(\alpha Z)$}
       &  {$P^{(3|2)}_{\rm tr1 }(\alpha Z)$}
       &  {$P^{(3|2)}_{\rm tr2 }(\alpha Z)$}
       &  {$P^{(3|2)}_{\rm     }(\alpha Z)$}  \\
        
\hline \\[-4mm]

  1  &  \multicolumn{1}{l}{\num[{scientific-notation=true,round-mode=places,round-precision=2}]{-0.000000000170}}  &    0.211964  &    0.000179  &    0.212143    \\ 
  2  &  \multicolumn{1}{l}{\num[{scientific-notation=true,round-mode=places,round-precision=2}]{-0.000000002571}}  &    0.215070  &    0.000379  &    0.215449    \\ 
  3  &  \multicolumn{1}{l}{\num[{scientific-notation=true,round-mode=places,round-precision=2}]{-0.000000012395}}  &    0.218186  &    0.000594  &    0.218781    \\ 
  4  &  \multicolumn{1}{l}{\num[{scientific-notation=true,round-mode=places,round-precision=2}]{-0.000000037471}}  &    0.221312  &    0.000820  &    0.222132    \\ 
  5  &  \multicolumn{1}{l}{\num[{scientific-notation=true,round-mode=places,round-precision=2}]{-0.000000087780}}  &    0.224446  &    0.001053  &    0.225499    \\ 
  6  &  \multicolumn{1}{l}{\num[{scientific-notation=true,round-mode=places,round-precision=2}]{-0.000000175078}}  &    0.227588  &    0.001291  &    0.228879    \\ 
  7  &  \multicolumn{1}{l}{\num[{scientific-notation=true,round-mode=places,round-precision=2}]{-0.000000312601}}  &    0.230737  &    0.001532  &    0.232268    \\ 
  8  &  \multicolumn{1}{l}{\num[{scientific-notation=true,round-mode=places,round-precision=2}]{-0.000000514817}}  &    0.233892  &    0.001773  &    0.235665    \\ 
  9  &  \multicolumn{1}{l}{\num[{scientific-notation=true,round-mode=places,round-precision=2}]{-0.000000797225}}  &    0.237054  &    0.002014  &    0.239067    \\ 
 10  &  \multicolumn{1}{l}{\num[{scientific-notation=true,round-mode=places,round-precision=2}]{-0.000001176184}}  &    0.240223  &    0.002251  &    0.242472    \\ 
 11  &  \multicolumn{1}{l}{\num[{scientific-notation=true,round-mode=places,round-precision=2}]{-0.000001668773}}  &    0.243398  &    0.002483  &    0.245879    \\ 
 12  &  \multicolumn{1}{l}{\num[{scientific-notation=true,round-mode=places,round-precision=2}]{-0.000002292669}}  &    0.246579  &    0.002710  &    0.249286    \\ 
 13  &  \multicolumn{1}{l}{\num[{scientific-notation=true,round-mode=places,round-precision=2}]{-0.000003066045}}  &    0.249767  &    0.002928  &    0.252692    \\ 
 14  &  \multicolumn{1}{l}{\num[{scientific-notation=true,round-mode=places,round-precision=2}]{-0.000004007485}}  &    0.252962  &    0.003137  &    0.256094    \\ 
 15  &  \multicolumn{1}{l}{\num[{scientific-notation=true,round-mode=places,round-precision=2}]{-0.000005135908}}  &    0.256163  &    0.003335  &    0.259493    \\ 
 16  &  \multicolumn{1}{l}{\num[{scientific-notation=true,round-mode=places,round-precision=2}]{-0.000006470507}}  &    0.259371  &    0.003521  &    0.262885    \\ 
 17  &  \multicolumn{1}{l}{\num[{scientific-notation=true,round-mode=places,round-precision=2}]{-0.000008030697}}  &    0.262587  &    0.003692  &    0.266271    \\ 
 18  &  \multicolumn{1}{l}{\num[{scientific-notation=true,round-mode=places,round-precision=2}]{-0.000009836072}}  &    0.265809  &    0.003849  &    0.269649    \\ 
 19  &  \multicolumn{1}{l}{\num[{scientific-notation=true,round-mode=places,round-precision=2}]{-0.000011906362}}  &    0.269039  &    0.003990  &    0.273017    \\ 
 20  &  \multicolumn{1}{l}{\num[{scientific-notation=true,round-mode=places,round-precision=2}]{-0.000014261408}}  &    0.272277  &    0.004112  &    0.276375    \\ 

\hline
\hline

\end{tabular}%

}


\end{minipage}
\end{table*}

\begin{figure*}[h!]
\begin{minipage}[b]{.485\textwidth}
  \includegraphics[width=\linewidth]{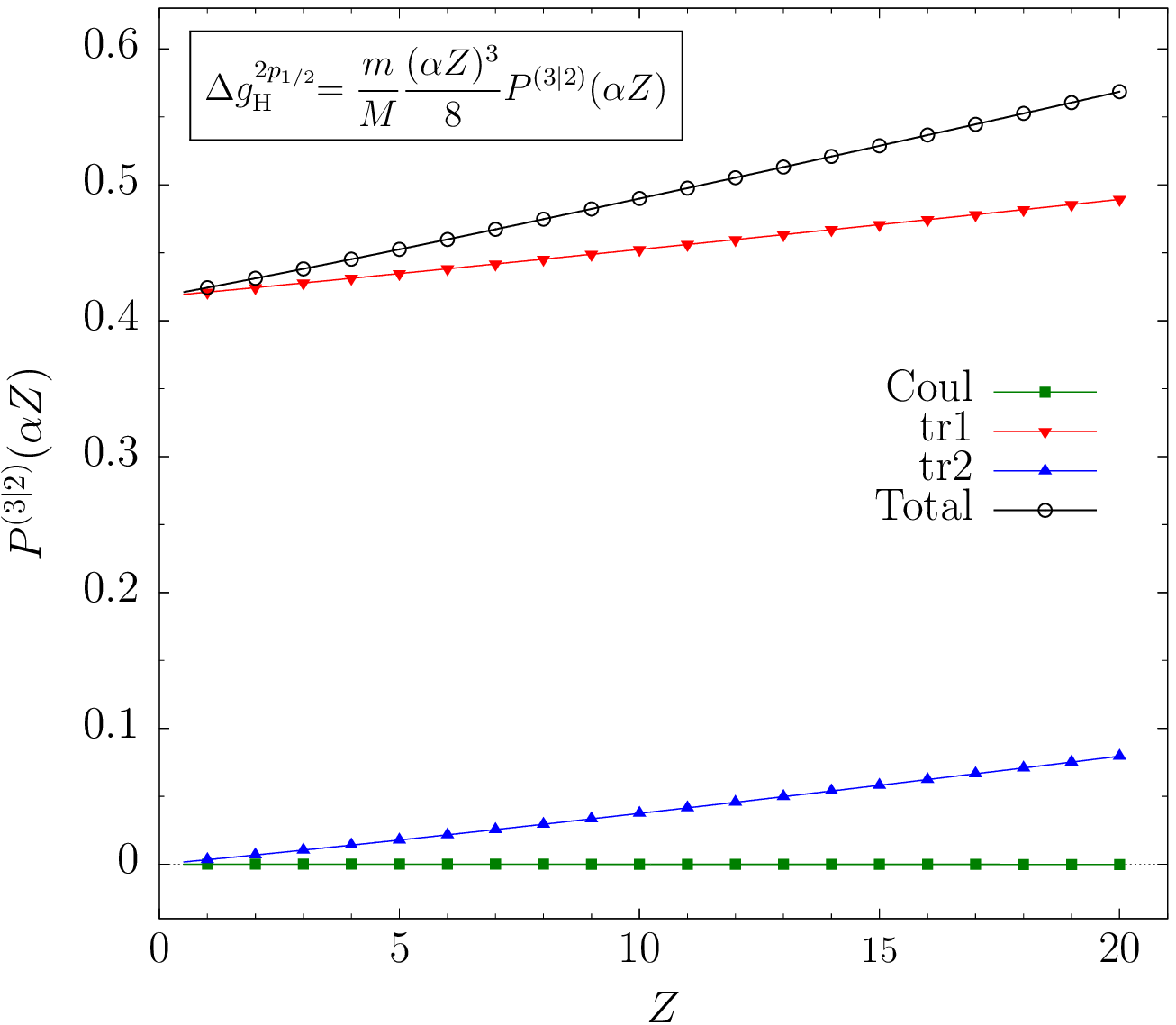}
  \caption{\label{fig:2p1} The Coulomb, one-transverse-photon, and two-transverse-photon contributions to the higher-order nuclear recoil effect on the $2p_{1/2}$ $g$ factor. The results are presented in terms of the function $P^{(3|2)}(\alpha Z)$ defined by Eq.~(\ref{eq:P}).}
\end{minipage}
\hfill
\begin{minipage}[b]{.485\textwidth}
  \includegraphics[width=\linewidth]{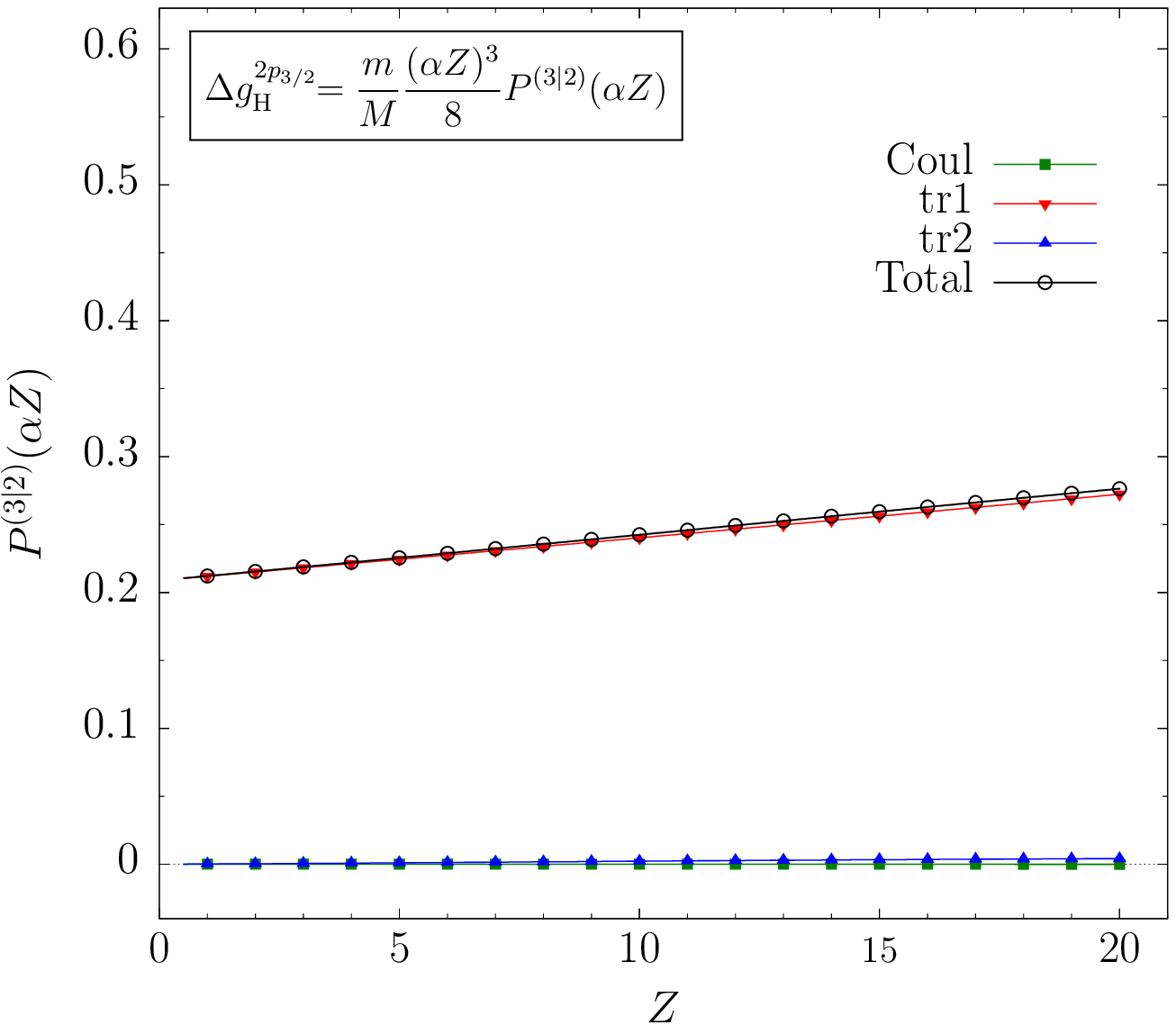}
  \caption{\label{fig:2p3} The Coulomb, one-transverse-photon, and two-transverse-photon contributions to the higher-order nuclear recoil effect on the $2p_{3/2}$ $g$ factor. The results are presented in terms of the function $P^{(3|2)}(\alpha Z)$ defined by Eq.~(\ref{eq:P}).}
\end{minipage}
\end{figure*}


The QED recoil contributions to the $g$ factors of the $2p_{1/2}$ and $2p_{3/2}$ states are given in Tables~\ref{tab:g_rec_2p1} and \ref{tab:g_rec_2p3}, respectively. For illustrative purposes, the results obtained are also plotted in Figs.~\ref{fig:2p1} and \ref{fig:2p3}. We note that for the $p$ states the $\Delta g_{\rm H}$ contribution possesses the $(\alpha Z)^3$ behavior in contrast to the $(\alpha Z)^5$ behavior found for the $s$ states. This fact is apparently related to the existence of the nonzero nonrelativistic limit for $\Delta g_{\rm L}^{2p_j}$ in Eqs.~(\ref{eq:g_L_2p1_expan}) and (\ref{eq:g_L_2p3_expan}) whereas the low-order contributions $\Delta g_{\rm L}^{1s}$ and $\Delta g_{\rm L}^{2s}$ in Eqs.~(\ref{eq:g_L_1s_expan}) and (\ref{eq:g_L_2s_expan}) are of the pure relativistic origin. For these reasons, the results in Tables~\ref{tab:g_rec_2p1} and \ref{tab:g_rec_2p3} and in Figs.~\ref{fig:2p1} and \ref{fig:2p3} are expressed in terms of the function $P^{(3|2)}(\alpha Z)$. From these data, one can conclude that for small $Z$ the higher-order part of the nuclear recoil effect for the $2p_{1/2}$ and $2p_{3/2}$ $g$ states is determined mainly by the one-transverse-photon contribution. The two-transverse-photon contribution is of the next order in $\alpha Z$ while the Coulomb contribution is almost negligible. 

Evaluating the limits of the QED recoil contributions to the $g$ factors of the $2p_{1/2}$ and $2p_{3/2}$ states at $\alpha Z \rightarrow 0$, we obtain 
\begin{align}
\label{eq:P_p_lim}
P^{(3|2)}_{2p_{1/2}}(0) = 0.417\,74(5) \, , 
\quad
P^{(3|2)}_{2p_{3/2}}(0)=0.208\,87(3) \, .
\end{align}
Based on Eqs.~(\ref{eq:g_L_2p1_expan}), (\ref{eq:g_L_2p3_expan}), and (\ref{eq:P_p_lim}), we note that the ratio of the QED recoil contributions to the $g$ factor of the $p$ states coincides with the analogous ratio for the low-order parts in the $\alpha Z\rightarrow 0$ limit,
\begin{align}
\lim_{\alpha Z\rightarrow 0} \frac{\Delta g_{\rm L}^{2p_{1/2}}}{\Delta g_{\rm L}^{2p_{3/2}}} 
=
\lim_{\alpha Z\rightarrow 0} \frac{\Delta g_{\rm H}^{2p_{1/2}}}{\Delta g_{\rm H}^{2p_{3/2}}} 
=
2 \, .
\end{align}

In the recent experiment~\cite{Arapoglou:2019:253001}, the ground-state $g$ factor of $^{40}{\rm Ar}^{13+}$ was measured to an accuracy of $10^{-9}$. The higher-order QED term evaluated in this paper amounts to $\Delta g_{\rm H}[^{40}{\rm Ar}^{13+}] = 2.1\cdot 10^{-9}$. This contribution, which is two times larger than the to-date experimental uncertainty, has to be taken into account, provided the many-electron QED and recoil corrections are evaluated to the required accuracy~\cite{Arapoglou:2019:253001}.


In addition, the theoretical value of the isotope shift in the atomic $g$ factor is determined mainly by the nuclear recoil and nuclear size effects. The measurement of the isotope difference of the bound-electron $g$ factor in lithiumlike calcium~\cite{Kohler:2016:10246} and the corresponding theoretical calculation~\cite{Shabaev:2017:263001} being in good agreement with each other pave the way for QED tests beyond the Furry picture at the strong coupling regime. In this regard, the high-precision measurements of the isotope shift of the bound-electron $g$ factor in boronlike ions are highly anticipated since the isotope dependence of the Zeeman effect can be evaluated to a very high accuracy exceeding significantly the accuracy of the $g$-factor calculations.

\section{Conclusion}

To summarize, in this paper we have evaluated the nuclear recoil effect of first order in $m/M$ on the bound-electron $g$ factors of the $n=1$ and $n=2$ states in H-like ions in the range $Z=1-20$. The calculations are performed to all orders in $\alpha Z$ by employing the fully relativistic approach. The numerical analysis of the behavior of the nuclear recoil contributions as functions of $Z$ is conducted. As the result, the most accurate theoretical predictions of the first-order in $m/M$ nuclear recoil effect on the bound-electron $g$ factor in hydrogenlike ions are obtained. The performed study is in demand in connection with the forthcoming experiments at the HITRAP/FAIR in Darmstadt and at the MPIK in Heidelberg.

\section{Acknowledgments}

This work was supported by the Russian Science Foundation (Grant No. 17-12-01097).



\begin{thebibliography}{10}

\bibitem{Shabaev:2015:031205}
V.~M.{~}Shabaev, D.~A.{~}Glazov, G.{~}Plunien, and A.~V.{~}Volotka,
\newblock J. Phys. Chem. Ref. Data {\bf 44},~031205 (2015).

\bibitem{Sturm:2017:4}
S.{~}Sturm, M.{~}Vogel, F.{~}{K{\"o}hler-Langes}, W.{~}Quint, K.{~}Blaum, and
  G.{~}Werth,
\newblock Atoms {\bf 5},~4 (2017).

\bibitem{Haffner:2000:5308}
H.{~}H{\"a}ffner, T.{~}Beier, N.{~}Hermanspahn, H.-J.{~}Kluge, W.{~}Quint,
  S.{~}Stahl, J.{~}Verd{\'u}, and G.{~}Werth,
\newblock Phys. Rev. Lett. {\bf 85},~5308 (2000).

\bibitem{Verdu:2004:093002}
J.{~}Verd{\'u}, S.{~}Djeki{\'c}, S.{~}Stahl, T.{~}Valenzuela, M.{~}Vogel,
  G.{~}Werth, T.{~}Beier, H.-J.{~}Kluge, and W.{~}Quint,
\newblock Phys. Rev. Lett. {\bf 92},~093002 (2004).

\bibitem{Sturm:2011:023002}
S.{~}Sturm, A.{~}Wagner, B.{~}Schabinger, J.{~}Zatorski, Z.{~}Harman,
  W.{~}Quint, G.{~}Werth, C.~H.{~}Keitel, and K.{~}Blaum,
\newblock Phys. Rev. Lett. {\bf 107},~023002 (2011).

\bibitem{Sturm:2013:030501_R}
S.{~}Sturm, A.{~}Wagner, M.{~}Kretzschmar, W.{~}Quint, G.{~}Werth, and
  K.{~}Blaum,
\newblock Phys. Rev. A {\bf 87},~030501(R) (2013).

\bibitem{Sturm:2014:467}
S.{~}Sturm, F.{~}K{\"o}hler, J.{~}Zatorski, A.{~}Wagner, Z.{~}Harman,
  G.{~}Werth, W.{~}Quint, C.~H.{~}Keitel, and K.{~}Blaum,
\newblock Nature {\bf 506},~467 (2014).

\bibitem{Zatorski:2017:012502}
J.{~}Zatorski, B.{~}Sikora, S.~G.{~}Karshenboim, S.{~}Sturm,
  F.{~}{K{\"o}hler-Langes}, K.{~}Blaum, C.~H.{~}Keitel, and Z.{~}Harman,
\newblock Phys. Rev. A {\bf 96},~012502 (2017).

\bibitem{Wagner:2013:033003}
A.{~}Wagner, S.{~}Sturm, F.{~}K{\"o}hler, D.~A.{~}Glazov, A.~V.{~}Volotka,
  G.{~}Plunien, W.{~}Quint, G.{~}Werth, V.~M.{~}Shabaev, and K.{~}Blaum,
\newblock Phys. Rev. Lett. {\bf 110},~033003 (2013).

\bibitem{Volotka:2014:253004}
A.~V.{~}Volotka, D.~A.{~}Glazov, V.~M.{~}Shabaev, I.~I.{~}Tupitsyn, and
  G.{~}Plunien,
\newblock Phys. Rev. Lett. {\bf 112},~253004 (2014).

\bibitem{Kohler:2016:10246}
F.{~}K{\"o}hler, K.{~}Blaum, M.{~}Block, S.{~}Chenmarev, S.{~}Eliseev,
  D.~A.{~}Glazov, M.{~}Goncharov, J.{~}Hou, A.{~}Kracke, D.~A.{~}Nesterenko,
  Y.~N.{~}Novikov, W.{~}Quint, E.~M.{~}Ramirez, V.~M.{~}Shabaev, S.{~}Sturm,
  A.~V.{~}Volotka, and G.{~}Werth,
\newblock Nat. Commun. {\bf 7},~10246 (2016).

\bibitem{Yerokhin:2017:062511}
V.~A.{~}Yerokhin, K.{~}Pachucki, M.{~}Puchalski, Z.{~}Harman, and
  C.~H.{~}Keitel,
\newblock Phys. Rev. A {\bf 95},~062511 (2017).

\bibitem{Glazov:2019:173001}
D.~A.{~}Glazov, F.{~}{K{\"o}hler-Langes}, A.~V.{~}Volotka, K.{~}Blaum,
  F.{~}Hei{\ss}e, G.{~}Plunien, W.{~}Quint, S.{~}Rau, V.~M.{~}Shabaev,
  S.{~}Sturm, and G.{~}Werth,
\newblock Phys. Rev. Lett. {\bf 123},~173001 (2019).

\bibitem{Arapoglou:2019:253001}
I.{~}Arapoglou, A.{~}Egl, M.{~}H{\"o}cker, T.{~}Sailer, B.{~}Tu, A.{~}Weigel,
  R.{~}Wolf, H.{~}Cakir, V.~A.{~}Yerokhin, N.~S.{~}Oreshkina,
  V.~A.{~}Agababaev, A.~V.{~}Volotka, D.~V.{~}Zinenko, D.~A.{~}Glazov,
  Z.{~}Harman, C.~H.{~}Keitel, S.{~}Sturm, and K.{~}Blaum,
\newblock Phys. Rev. Lett. {\bf 122},~253001 (2019).

\bibitem{Shabaev:2006:253002}
V.~M.{~}Shabaev, D.~A.{~}Glazov, N.~S.{~}Oreshkina, A.~V.{~}Volotka,
  G.{~}Plunien, H.-J.{~}Kluge, and W.{~}Quint,
\newblock Phys. Rev. Lett. {\bf 96},~253002 (2006).

\bibitem{Volotka:2014:023002}
A.~V.{~}Volotka and G.{~}Plunien,
\newblock Phys. Rev. Lett. {\bf 113},~023002 (2014).

\bibitem{Yerokhin:2016:100801}
V.~A.{~}Yerokhin, E.{~}Berseneva, Z.{~}Harman, I.~I.{~}Tupitsyn, and
  C.~H.{~}Keitel,
\newblock Phys. Rev. Lett. {\bf 116},~100801 (2016).

\bibitem{Furry:1951:115}
W.~H.{~}Furry,
\newblock Phys. Rev. {\bf 81},~115 (1951).

\bibitem{Shabaev:2002:091801}
V.~M.{~}Shabaev and V.~A.{~}Yerokhin,
\newblock Phys. Rev. Lett. {\bf 88},~091801 (2002).

\bibitem{Shabaev:2001:052104}
V.~M.{~}Shabaev,
\newblock Phys. Rev. A {\bf 64},~052104 (2001).

\bibitem{Shabaev:2017:263001}
V.~M.{~}Shabaev, D.~A.{~}Glazov, A.~V.{~}Malyshev, and I.~I.{~}Tupitsyn,
\newblock Phys. Rev. Lett. {\bf 119},~263001 (2017).

\bibitem{Shabaev:2018:032512}
V.~M.{~}Shabaev, D.~A.{~}Glazov, A.~V.{~}Malyshev, and I.~I.{~}Tupitsyn,
\newblock Phys. Rev. A {\bf 98},~032512 (2018).

\bibitem{Malyshev:2017:765}
A.~V.{~}Malyshev, V.~M.{~}Shabaev, D.~A.{~}Glazov, and I.~I.{~}Tupitsyn,
\newblock JETP Lett. {\bf 106},~765 (2017).

\bibitem{Shchepetnov:2015:012001}
A.~A.{~}Shchepetnov, D.~A.{~}Glazov, A.~V.{~}Volotka, V.~M.{~}Shabaev,
  I.~I.{~}Tupitsyn, and G.{~}Plunien,
\newblock J. Phys.: Conf. Ser. {\bf 583},~012001 (2015).

\bibitem{Glazov:2018:457}
D.~A.{~}Glazov, A.~V.{~}Malyshev, V.~M.{~}Shabaev, and I.~I.{~}Tupitsyn,
\newblock Opt. Spectrosc. {\bf 124},~457 (2018).

\bibitem{Aleksandrov:2018:062521}
I.~A.{~}Aleksandrov, D.~A.{~}Glazov, A.~V.{~}Malyshev, V.~M.{~}Shabaev, and
  I.~I.{~}Tupitsyn,
\newblock Phys. Rev. A {\bf 98},~062521 (2018).

\bibitem{Lindenfels:2013:023412}
D.{~}{von Lindenfels}, M.{~}Wiesel, D.~A.{~}Glazov, A.~V.{~}Volotka,
  M.~M.{~}Sokolov, V.~M.{~}Shabaev, G.{~}Plunien, W.{~}Quint, G.{~}Birkl,
  A.{~}Martin, and M.{~}Vogel,
\newblock Phys. Rev. A {\bf 87},~023412 (2013).

\bibitem{Vogel:2019:1800211}
M.{~}Vogel, M.~S.{~}Ebrahimi, Z.{~}Guo, A.{~}Khodaparast, G.{~}Birkl, and
  W.{~}Quint,
\newblock Ann. Phys. (Berlin) {\bf 531},~1800211 (2019).

\bibitem{Sturm:2019:1425}
S.{~}Sturm, I.{~}Arapoglou, A.{~}Egl, M.{~}H{\"o}cker, S.{~}Kraemer,
  T.{~}Sailer, B.{~}Tu, A.{~}Weigel, R.{~}Wolf, J.~C.{~}{L{\'o}pez-Urrutia},
  and K.{~}Blaum,
\newblock Eur. Phys. J. Spec. Top. {\bf 227},~1425 (2019).

\bibitem{Shabaev:2002:062104}
V.~M.{~}Shabaev, D.~A.{~}Glazov, M.~B.{~}Shabaeva, V.~A.{~}Yerokhin,
  G.{~}Plunien, and G.{~}Soff,
\newblock Phys. Rev. A {\bf 65},~062104 (2002).

\bibitem{Shabaev:1991:4479}
V.~M.{~}Shabaev,
\newblock J. Phys. B: At. Mol. Opt. Phys. {\bf 24},~4479 (1991).

\bibitem{Johnson:1988:307}
W.~R.{~}Johnson, S.~A.{~}Blundell, and J.{~}Sapirstein,
\newblock Phys. Rev. A {\bf 37},~307 (1988).

\bibitem{Sapirstein:1996:5213}
J.{~}Sapirstein and W.~R.{~}Johnson,
\newblock J. Phys. B: At. Mol. Opt. Phys. {\bf 29},~5213 (1996).

\bibitem{Yerokhin:2015:033103}
V.~A.{~}Yerokhin and V.~M.{~}Shabaev,
\newblock J. Phys. Chem. Ref. Data {\bf 44},~033103 (2015).

\end{thebibliography}


\end{document}